\documentclass[aps,pre,twocolumn,floatfix]{revtex4}

\usepackage{graphicx}
\usepackage{amsmath}
\usepackage{amssymb}

\newcommand{\be}{\begin{equation}}
\newcommand{\ee}{\end{equation}}
\newcommand{\ba}{\begin{array}{l}}
\newcommand{\ea}{\end{array}}
\newcommand{\re}[1]{(\ref{#1})}
\newcommand{\ci}[1]{\cite{#1}}
\newcommand{\banonum}{\begin{eqnarray*}}
\newcommand{\eanonum}{\end{eqnarray*}}
\newcommand{\baa}{\begin{eqnarray}}
\newcommand{\eaa}{\end{eqnarray}}
\newcommand{\bfr}{\begin{flushright}}
\newcommand{\efr}{\end{flushright}}
\newcommand{\bfl}{\begin{flushleft}}
\newcommand{\efl}{\end{flushleft}}
\begin{document}

\title{Discrete sine-Gordon equation on metric graphs:\\ A simple model for Josephson junction networks}
\author{M.E. Akramov$^{1}$, J.R. Yusupov$^{2}$, I.N. Askerzade$^3$ and D.U. Matrasulov$^{4}$}
\affiliation{$^1$National University of Uzbekistan, 4 Universitet Str., 100174, Tashkent, Uzbekistan\\
$^2$Kimyo International University in Tashkent, 156 Usman Nasyr Str., 100121, Tashkent, Uzbekistan\\
$^3$Department of Computer Engineering, Ankara University, 06100, Ankara, Turkey\\
$^4$Turin Polytechnic University in Tashkent, 17 Niyazov Str., 100095, Tashkent, Uzbekistan}

\begin{abstract}
We consider discrete sine-Gordon equation on branched domains. The latter is modeled in terms of the metric graphs with discrete bonds having the form of the branched 1D chains. Exact analytical solutions of the problem are obtained for special case of the constraints given by in terms of simple sum rule. Numerical solution is obtained when the constraint is not fulfilled. 
\end{abstract}

\maketitle

\section{Introduction}
Sine-Gordon solitons attracted much attention in different contexts, from solid state mechanics to quantum field theory and Josepshon junctions (see, Refs.\ci{Orfandis}-\ci{Caputo2}). Recently, special attention was given to the study of sine-Gordon solitons in branched domains and networks.
Particle and wave dynamics in low-dimensional branched domains is of importance for many  practically important problems arising in newly emerging technologies. As many device structures and functional materials have branched structure or networked form, controlling of wave propagation in such structures play crucial role for device optimization and material design. Solving such a task requires developing realistic and physically acceptable models of the wave dynamics in such systems. 
An effective way for modelling of solitons in networks can be based on the solution of different nonlinear evolution equations (approving soliton solutions) on metric graphs. Such a task has become subject of extensive study recently (see, Refs. \ci{Hadi2005}-\ci{BJJEPL}).
Especially, this concerns condensed matter systems, where particle and wave transport in linear(quantum) and nonlinear regimes, where the problem of tunable transport is of crucial importance. Such tasks appear, e.g., in BEC dynamics, optical harmonic generation low-dimensional quantum materials, Josephson junctions, etc. Tuning the wave propagation process in these structures allows optimization of material functional properties and improving of the device performance.  In this paper we consider the model, which is directly related to the problem  soliton dynamics in Josephson junction networks. Namely, we consider sine-Gordon equation on discrete branched lattice. Sine-Gordon equation in such lattice can describe Josephson junction(JJ) network consisting of branched JJ arrays, where superconducting leads are separated by point-like insulators or normal metals. Different versions of such branched JJ-arrays have been considered earlier in the Refs.\ci{Sodano1}-\ci{Ovchin1}. However, these studies did not consider soliton dynamics in such structures and did not use metric grapgh based approach for sine-Gordon equation. In such lattice, the phase difference (on each junction) between the leads is described in terms of the discrete sine-Gordon equation. Imposing the boundary conditions in the of weight continuity and Kirchhoff rules at the branching point, we derive constraints ensuring integrability of the Discrete sine-Gordon equation in metric graph. Such constraint can be written in the form of simple sum rule in terms of the nonlinearity coefficients. Apart from the branched Josephson junction arrays,  within the Frenkel-Kontorova model \ci{kivshar98,Kivsharbook}, discrete sine-Gordon equation on matric graphs can be used for modeling deformation propagation in branched solid materials. In both cases, the main problem having practical interest is tunable propagation of sine-Gordon soliton in a branched structure. Here we show that in case when the problem is integrable and sine-Gordon solitons pass through the graph vertex without reflection, i.e. no backscattering at the branching points for integrable case. The paper is organized as follows/ In the next section we briefly recall the discrete sine-Gordon equation on a line. In section III we present formulation of the task and its solution for star branched graph. Section IV extends the study for the case of loop graph. Finally, the section V provides some concluding remarks.

\section{Discrete sine-Gordon equation on a line}
The discrete sine-Gordon (DSG) equation  follows from the Hamiltonian of the Frenkel-Kontorova model which is is given as ~\cite{kivshar98}
\begin{multline}
H=\sum_{n=-\infty}^{+\infty} \bigg[  \frac{1}{2}\bigg( \frac{du_n}{dt} \bigg)^2+
\beta(1-\cos{u_n})\\
+\frac{a}{2}(u_{n+1}-u_{n})^2
\bigg].
\end{multline}
Equation of motion for this  Hamiltonian, leads to the following standard discretized sine-Gordon equation:
\begin{equation}
\frac{d^2 u_n}{dt^2}-a(u_{n+1}-2u_n+u_{n-1})+\beta\sin{u_n}=0,\label{line}
\end{equation}
where $a$ and $\beta$ are constant coefficients.
For the above discrete sine-Gordon (DSG) equation one can write the charge in the form \ci{Kivsharbook,Panosbook}

\begin{equation}
Q=\frac{1}{2\pi} \sqrt{\frac{a}{\beta}} \sum_{n=-\infty}^{+\infty} (u_{n+1}-u_{n}). \label{charge_line}
\end{equation}

The other conservative quantities are the energy given by Eq.~(1) and momentum, which is given by
\begin{eqnarray}
    P=\frac{\sqrt{a}}{\beta} \sum_{n=-\infty}^{+\infty} \frac{du_n}{dt}(u_{n+1}-u_n). \label{momentum}
\end{eqnarray}
Kink soliton solution of Eq.\re{line} can be written as (for $\alpha =1,$ and $\beta =1$)
The solution of Eq.\eqref{dsg_line} in the form
\begin{equation}
v_n(t)=4\arctan\left[ \exp\left(\pm \frac{n-n_0-v t}{\sqrt{1-v^2}}\right) \right].\label{sol01}
\end{equation}
In the next section we use this solution to construct soliton solution of DSGE on a graph.

\section{Discrete sine-Gordon equation on a star graph}
\begin{figure}[t!]
\includegraphics[width=80mm]{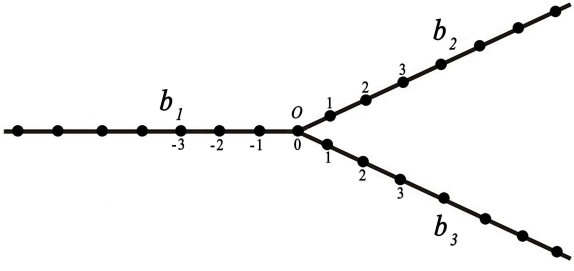}
\caption{Basic star graph.} \label{fig:star}
\end{figure}

Consider the star graph that consists of three semi-infinite chains connected at the vertex (see Fig. \ref{fig:star}). For the first $j=1$ bond $n$ is numbered as $n \in b_{1}=\{-1,-2,-3...\}$, where $(j,-1)$ stands for the point nearest to the vertex. 
For the right handed $j=2,3$ bonds $n$ is numbered as $n \in b_j=\{0,1,2,...\}$, where $(j,0)$ means the branching point.

The DSG equation is written on the each bond of the star graph as follows
\begin{equation}
\frac{d^2 u_{j,n}}{dt^2}-a_j(u_{j,n+1}-2u_{j,n}+u_{j,n-1})+\beta_j\sin{u_{j,n}}=0.\label{dsg_star}
\end{equation}
The charge for Eq. \eqref{dsg_star} can be written as
\begin{equation}
Q=\frac{1}{2\pi} \sum_{j=1}^3 \sqrt{\frac{a_j}{\beta_j}} \sum_{b_j} (u_{j,n+1}-u_{j,n}). \label{charge_star}
\end{equation}
Assuming the following relation at the virtual site, $(1,0)$:  $u_{1,0}=\alpha_2 u_{2,0}+\alpha_3 u_{3,0}$  ($\alpha_2$ and $\alpha_3$ constant coefficients), from the charge conservation law  which is given as
\begin{equation}
\frac{dQ}{dt}=\frac{1}{2\pi} \left[ 
\sqrt{\frac{a_1}{\beta_1}} \frac{du_{1,0}}{dt} - 
\sqrt{\frac{a_2}{\beta_2}} \frac{du_{2,0}}{dt} -
\sqrt{\frac{a_3}{\beta_3}} \frac{du_{3,0}}{dt} \right] =0, \label{charge01}
\end{equation}
we have 
\begin{equation}
\sqrt{\frac{a_1}{\beta_1}} \frac{du_{1,0}}{dt}=
\sqrt{\frac{a_2}{\beta_2}} \frac{du_{2,0}}{dt}+
\sqrt{\frac{a_3}{\beta_3}} \frac{du_{3,0}}{dt}. \label{charge02}
\end{equation}

Similarly, for the energy, which is given by 
\begin{multline}
E=\sum_{j=1}^3\frac{1}{\beta_j}\sum_{b_j} \bigg[  \frac{1}{2}\bigg( \frac{du_{j,n}}{dt} \bigg)^2+
\beta_j(1-\cos{u_{j,n}})\\
+\frac{a_j}{2}(u_{j,n+1}-u_{j,n})^2
\bigg],
\end{multline}
we have the following conservation law
\begin{multline}
\frac{dE}{dt}=\frac{a_1}{\beta_1}\frac{du_{1,0}}{dt}(u_{1,0}-u_{1,-1})
-\frac{a_2}{\beta_2}\frac{du_{2,0}}{dt}(u_{2,0}-u_{2,-1})\nonumber\\-
\frac{a_3}{\beta_3}\frac{du_{3,0}}{dt}(u_{3,0}-u_{3,-1})=0, \label{energy}
\end{multline}
that leads to the following condition:
\begin{gather}
\frac{a_1}{\beta_1}\frac{du_{1,0}}{dt}(u_{1,0}-u_{1,-1})=
\frac{a_2}{\beta_2}\frac{du_{2,0}}{dt}(u_{2,0}-u_{2,-1})\nonumber\\+
\frac{a_3}{\beta_3}\frac{du_{3,0}}{dt}(u_{3,0}-u_{3,-1}). \label{energy01}
\end{gather}
\begin{figure}[t!]
\includegraphics[width=80mm]{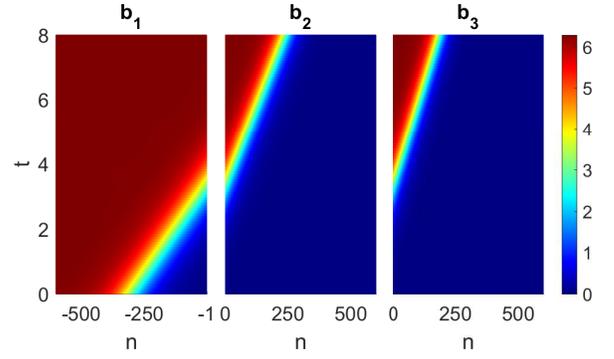}
\caption{Contour plot of the numerical solution of DSG equation on the star graph when the sum rule \eqref{contr1} is fulfilled, i.e., the parameters are chosen as $\beta_1=\beta_2=\beta_3=1, \; a_1=2.56, \; a_2=0.81, \; a_3=0.49$} \label{fig:contour_fulfilled}
\end{figure}

From Eqs. \eqref{charge02} and \eqref{energy01} we can obtain  the following "quasi" vertex boundary conditions:
\begin{gather}
\sqrt{\frac{a_1}{\beta_1}} u_{1,0}=
\sqrt{\frac{a_2}{\beta_2}} u_{2,0}+
\sqrt{\frac{a_3}{\beta_3}} u_{3,0},\label{vbc01}\\
\sqrt{\frac{a_1}{\beta_1}} (u_{1,0}-u_{1,-1})=
\sqrt{\frac{a_2}{\beta_2}} (u_{2,0}-u_{2,-1})\nonumber\\
=\sqrt{\frac{a_3}{\beta_3}} (u_{3,0}-u_{3,-1}).\label{vbc02}
\end{gather}
From Eqs.\eqref{vbc01} and \eqref{vbc02} one can define $u_{j,n}$ at the virtual $(j,n)=\{(1,0),(2,-1),(3,-1)\}$ sites:
\begin{gather}
u_{1,0}=\sqrt{\frac{\beta_1}{a_1}}\left(   \sqrt{\frac{a_2}{\beta_2}}u_{2,0}+\sqrt{\frac{a_3}{\beta_3}}u_{3,0}    \right),\nonumber\\
u_{2,-1}=\sqrt{\frac{\beta_2}{a_2}}\left(   \sqrt{\frac{a_1}{\beta_1}}u_{1,-1}-\sqrt{\frac{a_3}{\beta_3}}u_{3,0}    \right),\\
u_{3,-1}=\sqrt{\frac{\beta_3}{a_3}}\left(   \sqrt{\frac{a_1}{\beta_1}}u_{1,-1}-\sqrt{\frac{a_2}{\beta_2}}u_{2,0}    \right).\nonumber
\end{gather}

Furthermore, we assume that $\alpha_j$ and $\beta_k$ fulfill the following relations:
\begin{eqnarray}
\beta_1=\beta_2=\beta_3,\quad \sqrt{a_1}=\sqrt{a_2}+\sqrt{a_3}. \label{contr1}
\end{eqnarray}

Then the soliton (kink) solution of Eq.\eqref{dsg_star} can be written in terms of solution of Eq.\eqref{line} as
\begin{equation}
u_{j,n}(t)=4\arctan\left[ \exp\left(\pm \frac{\sqrt{\frac{\beta_j}{a_j}} (n-n_0)-\beta_j v t}{\sqrt{1-v^2}}\right) \right].\label{sol02}
\end{equation}
\begin{figure}[t!]
\includegraphics[width=80mm]{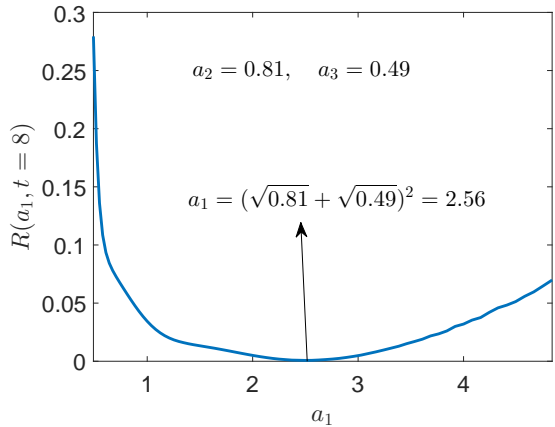}
\caption{The plot reflection coefficient versus $a_1$ which is determined as $R=E_1|_{t=8}/E_1|_{t=0}$.} \label{fig:R_fulfilled}
\end{figure}
\begin{figure}[t!]
\includegraphics[width=80mm]{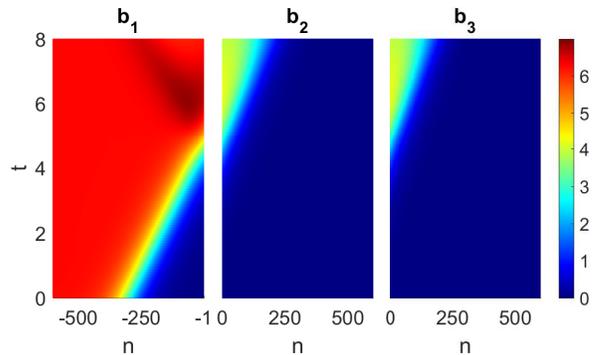}
\caption{Contour plot of the numerical solution of DSG equation on the star graph when the sum rule \eqref{contr1} is broken, i.e., the parameters are chosen as $\beta_1=\beta_2=\beta_3=1, \; a_1=1.44, \; a_2=1, \; a_3=0.81$} \label{fig:contour_broken}
\end{figure}

\begin{figure}[t!]
\includegraphics[width=80mm]{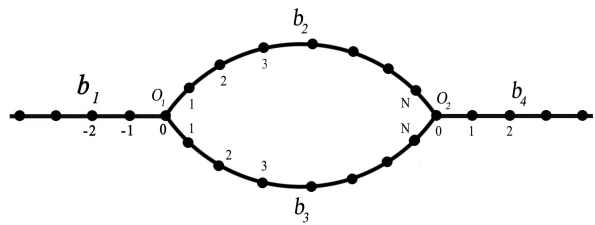}
\caption{The loop graph.} \label{fig:loop}
\end{figure}

It should be noted that Eq.\re{contr1} presents condition (constraint) for integrability of the problem given by Eqs.\re{dsg_star}, \re{vbc01} and \re{vbc02}. In other words, the discrete sine-Gordon equation \re{dsg_star} on metric star graph presented in Fig.1 is integrable if and only if sum rule in Eq.\re{contr1} fulfilled. In Fig. 2 sptio-temporal evolution of the sine-Gordon soliton  obtained using the exact solution given by Eq.\re{sol02} is plotted. One can clearly observe from this plot that transmission of the sine-Gordon soliton through the branching point of the graph is reflectionless, i.e., no backscattering is observed for the case, when the problem is integrable. This feature can be confirmed by direct computing reflection coefficient (as the ratio of soliton energy on bond 1 to the total one). 

In Fig. 3 the plot of reflection coefficient, $R = E_1|_{t=8}/E_1|_{t=0}$
on $a_3$, where $a_1 = 0.81$, $a_2 = 0.49$, is plotted. As it can be seen from this plot, at the value of $a_3 =2.51$, which corresponds to fulfilling of the sum rule, $R$ becomes zero.
For the case, when the problem is not integrable, i.e.. when the sum rule in Eq.\re{contr1} is broken, the problem need to be solved numerically. 
The plot of the solution for the case, when the sum rule is broken is presented in Fig.4. The plot shows that transmission of sine-Gordon soliton through the branching point is accompanies by reflection. 
\begin{figure}[t!]
\includegraphics[width=80mm]{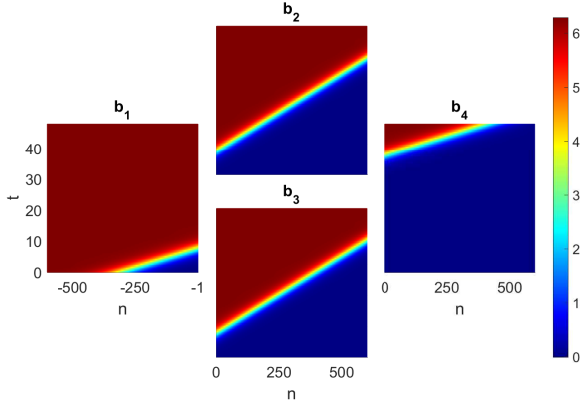}
\caption{Contour plot of the numerical solution of DSG equation on the loop graph when the sum rule \eqref{sum_rule_loop} is fulfilled, i.e., the parameters are chosen as $\beta_1=\beta_2=\beta_3=\beta_4=1, \; a_1=a_4=1, \; a_2=a_3=0.25$} \label{fig:contour_loop}
\end{figure}

\section{Extending to the loop graph}
The above treatment of the DSG equation on star graph can be extended to the case of other graph topologies, e.g., to loop graph presented in Fig. \ref{fig:loop}. The graph consists of two semi infinite and two finite chains connected to each other at two vertices. 
On each bond of the loop graph we have the DSG equation given by \eqref{dsg_star}.
The charge and the energy for such structure are given as (respectively)
\begin{gather}
Q=\frac{1}{2\pi} \sum_{j=1}^4 \sqrt{\frac{a_j}{\beta_j}} \sum_{b_j} (u_{j,n+1}-u_{j,n}), \label{charge_loop}\\
E=\sum_{j=1}^4\frac{1}{\beta_j}\sum_{b_j} \bigg[  \frac{1}{2}\bigg( \frac{du_{j,n}}{dt} \bigg)^2+
\beta_j(1-\cos{u_{j,n}})\nonumber\\
+\frac{a_j}{2}(u_{j,n+1}-u_{j,n})^2
\bigg]. \label{energy_loop}
\end{gather}

From the conservation laws for these quantities we obtain the following relations:
\begin{widetext}
\begin{gather}
\sqrt{\frac{a_1}{\beta_1}} \frac{du_{1,0}}{dt}+
\sqrt{\frac{a_2}{\beta_2}} \frac{du_{2,N+1}}{dt}+
\sqrt{\frac{a_3}{\beta_3}} \frac{du_{3,N+1}}{dt}=%\nonumber\\=
\sqrt{\frac{a_2}{\beta_2}} \frac{du_{2,0}}{dt}+
\sqrt{\frac{a_3}{\beta_3}} \frac{du_{3,0}}{dt}+
\sqrt{\frac{a_4}{\beta_4}} \frac{du_{4,0}}{dt}.\label{vbc_loop1}\\
\sqrt{\frac{a_1}{\beta_1}} (u_{1,0}-u_{1,-1})=
\sqrt{\frac{a_2}{\beta_2}} (u_{2,N+1}-u_{2,N})=%\nonumber\\
\sqrt{\frac{a_3}{\beta_3}} (u_{3,N+1}-u_{3,N})\nonumber\\=
\sqrt{\frac{a_2}{\beta_2}} (u_{2,0}-u_{1,-1})=
\sqrt{\frac{a_3}{\beta_3}} (u_{3,0}-u_{2,-1})=%\nonumber\\
\sqrt{\frac{a_4}{\beta_4}} (u_{4,0}-u_{3,-1}).
\label{vbc_loop2}
\end{gather}
\end{widetext}

From  Eqs. \eqref{vbc_loop1} and \eqref{vbc_loop2} one can obtain vertex quasi-boundary conditions, which can be written as
\begin{gather}
\sqrt{\frac{a_1}{\beta_1}} \frac{du_{1,0}}{dt}=
\sqrt{\frac{a_2}{\beta_2}} \frac{du_{2,0}}{dt}+
\sqrt{\frac{a_3}{\beta_3}} \frac{du_{3,0}}{dt},\nonumber\\
\sqrt{\frac{a_2}{\beta_2}} \frac{du_{2,N+1}}{dt}+
\sqrt{\frac{a_3}{\beta_3}} \frac{du_{3,N+1}}{dt}=
\sqrt{\frac{a_4}{\beta_4}} \frac{du_{4,0}}{dt}.
\label{vbc_loop3}
\end{gather}
\begin{gather}
\sqrt{\frac{a_1}{\beta_1}} (u_{1,0}-u_{1,-1})=
\sqrt{\frac{a_2}{\beta_2}} (u_{2,0}-u_{1,-1})\nonumber\\=
\sqrt{\frac{a_3}{\beta_3}} (u_{3,0}-u_{2,-1})\nonumber\\
\sqrt{\frac{a_2}{\beta_2}} (u_{2,N+1}-u_{2,N})=
\sqrt{\frac{a_3}{\beta_3}} (u_{3,N+1}-u_{3,N})\nonumber\\=
\sqrt{\frac{a_4}{\beta_4}} (u_{4,0}-u_{3,-1}).
\label{vbc_loop4}
\end{gather}
From Eqs. \eqref{vbc_loop3} and \eqref{vbc_loop4} one can define $u_{j,n}$ at the virtual
$(j,n)=\{(1,0),(2,-1),(3,-1)\}$ and 
$(j,n)=\{(2,N+1),(3,N+1),(4,-1)\}$ sites
\begin{gather}
u_{1,0}=\sqrt{\frac{\beta_1}{a_1}}\left(   \sqrt{\frac{a_2}{\beta_2}}u_{2,0}+\sqrt{\frac{a_3}{\beta_3}}u_{3,0}    \right),\nonumber\\
u_{2,-1}=\sqrt{\frac{\beta_2}{a_2}}\left(   \sqrt{\frac{a_1}{\beta_1}}u_{1,-1}-\sqrt{\frac{a_3}{\beta_3}}u_{3,0}    \right),\\
u_{3,-1}=\sqrt{\frac{\beta_3}{a_3}}\left(   \sqrt{\frac{a_1}{\beta_1}}u_{1,-1}-\sqrt{\frac{a_2}{\beta_2}}u_{2,0}    \right),\nonumber
\end{gather}
and
\begin{gather}
u_{2,N+1}=\sqrt{\frac{\beta_2}{a_2}}\left(   \sqrt{\frac{a_4}{\beta_4}}u_{4,0}-\sqrt{\frac{a_3}{\beta_3}}u_{3,N}    \right),\nonumber\\
u_{3,N+1}=\sqrt{\frac{\beta_3}{a_3}}\left(   \sqrt{\frac{a_4}{\beta_4}}u_{4,0}-\sqrt{\frac{a_2}{\beta_2}}u_{2,N}    \right),\\
u_{4,-1}=\sqrt{\frac{\beta_4}{a_4}}\left(   \sqrt{\frac{a_2}{\beta_2}}u_{2,N}+\sqrt{\frac{a_3}{\beta_3}}u_{3,N}    \right).\nonumber
\end{gather}

\begin{figure}[t!]
\includegraphics[width=80mm]{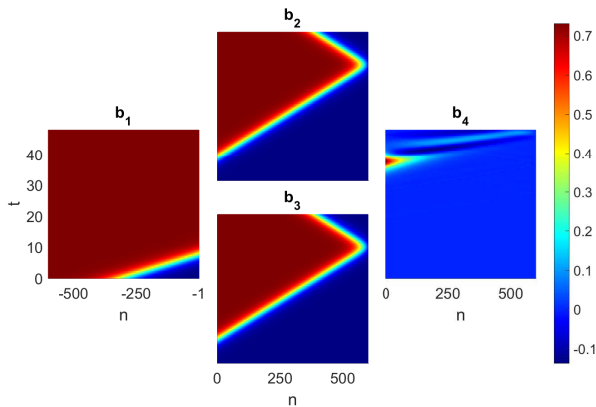}
\caption{Contour plot of the numerical solution of DSG equation on the loop graph when the sum rule \eqref{sum_rule_loop} is broken, i.e., the parameters are chosen as $\beta_1=\beta_2=\beta_3=\beta_4=1, \; a_1=1, \; a_2=a_3=0.25, \; a_4=1.1$} \label{fig:contour_loop}
\end{figure}

Again, as it was done in the case of star branched graph, by substituting the solution of DSGE on a line into the vertex quasi-boundary conditions given by Eqs. \eqref{vbc_loop3} and \eqref{vbc_loop4}, one can obtain the following constraints:
\begin{gather}
\beta_1=\beta_2=\beta_3=\beta_4, \quad \sqrt{a_1}=\sqrt{a_2}+\sqrt{a_3}, \quad a_4=a_1.
\label{sum_rule_loop}
\end{gather}
Fulfilling of this constraint implies that the problem given by Eqs.\re{dsg_star}, \eqref{vbc_loop3} and \eqref{vbc_loop4} is integrable and its (kink) soliton solution can be written as
\begin{multline}
u_{j,n}(t)=4\arctan\left[ 
\exp\left(
\pm \frac{\sqrt{\frac{\beta_j}{a_j}} n-n_0-\beta_j v t}{\sqrt{1-v^2}}
\right) 
\right], \\ j=1,2,3  \label{sol_loop1}
\end{multline}
and
\begin{gather}
u_{4,n}(t)=4\arctan\left[ 
\exp\left(
\pm \frac{\sqrt{\frac{\beta_4}{a_4}}(n+N+1)-n_0-\beta_4 v t}{\sqrt{1-v^2}}
\right) 
\right].\label{sol_loop2}
\end{gather}
In Fig. 6 the contour plot of the soliton solution for the case, when the sum rule in Eq. \re{sum_rule_loop} is fulfilled is presented. One can observe again absence of the backscattering in case, when the problem is integrable. Fig. 7 presents similar plots for non-integravble case, when the solutions are obtained numerically. Appearing of reflection in transmission of soliton through the branching point can be clearly seen from the plot.

\section{Conclusions}
We studied the problem of discrete sine-Gordon equation on a branched lattice by addressing the problem of integrability and soliton solutions. It is shown that the problem approves exact soliton solutions, provided the nonlinearity coupling constant (penetration depth in case of  Josephson junction) assigned to each bond of the graph fulfill certain sum rule. Such case associated also with the reflectionless transmission of sine-Gordon solitons through the branching point.  For the cases, when the sum rule is not fulfilled the problem solved analytically. It is shown for the latter case that reflection of soliton at the vertex can be observed. Although the above treatment dealt with the star and loop  graphs, the approach developed here can be directly applied for arbitrary branching topology. The above results can be directly applied for modeling of the dynamics of sine-Gordon solitons in branched arrays of Josephson junctions.
\section{Acknowledgements}
This work is supported by joint grant the Ministry of Innovative Development of Uzbekistan (Ref. Nr. MRT-2130213155) and TUBITAK (Ref. Nr. 221N123)

\end{document}